\documentclass{bioinfo}
\copyrightyear{2007}
\pubyear{2007}

\begin{document}
\firstpage{1}
\title[Short title]{Detection of the dominant direction of
information flow in densely interconnected regulatory networks}
\author[Ispolatov, I. and Maslov, S.]{I. Ispolatov\,$^{\rm a}$\footnote{Permanent address:
Departamento de Fisica, Universidad de Santiago de Chile,
Casilla 302, Correo 2, Santiago, Chile}, S. Maslov\,$^{\rm b}$
}
\address{$^{\rm a}$
Ariadne Genomics Inc., 9430 Key West Ave.  Suite 113
Rockville, MD 20850, USA, $^{\rm b}$ Department of Condensed Matter Physics
and Materials Science,
Brookhaven National Laboratory, Upton, New York 11973, USA}
\maketitle

\begin{abstract}

\section{Motivation:}
Finding the dominant direction of flow of
information in densely interconnected regulatory or
signaling networks is required in many applications in
computational biology and neuroscience.
This is achieved by first identifying and removing links which close up
feedback loops in the original network and hierarchically
arranging nodes in the remaining network. In mathematical language
this corresponds to a problem of making a graph acyclic
by removing as few links as possible and thus
altering the original graph in the least possible way. Practically in all
applications the exact solution of this problem requires an enumeration of all
combinations of removed links, which is computationally intractable.
\section{Results:}
We introduce and compare two algorithms: the deterministic,
'greedy' algorithm that preferentially cuts the links that participate in the
largest number of
feedback cycles, and the probabilistic one based on a simulated
annealing of a hierarchical layout of the network which minimizes
the number of ``backward'' links going from lower to higher hierarchical levels.
We find that the annealing algorithm outperforms the deterministic one in terms of
speed, memory requirement, and the actual number of removed links.
Implications  for system biology and directions for further research are
discussed.
\section{Availability:} Source codes of $F90$ and Matlab implementation of these
two algorithms are available from the authors upon request.
\section{Contact:} \href{slava@ariadnegenomics.com},
\href{maslov@bnl.gov}
\end{abstract}

\section{Introduction}
\begin{figure}[!tpb]
\centerline{\includegraphics[width=3in,angle=0]{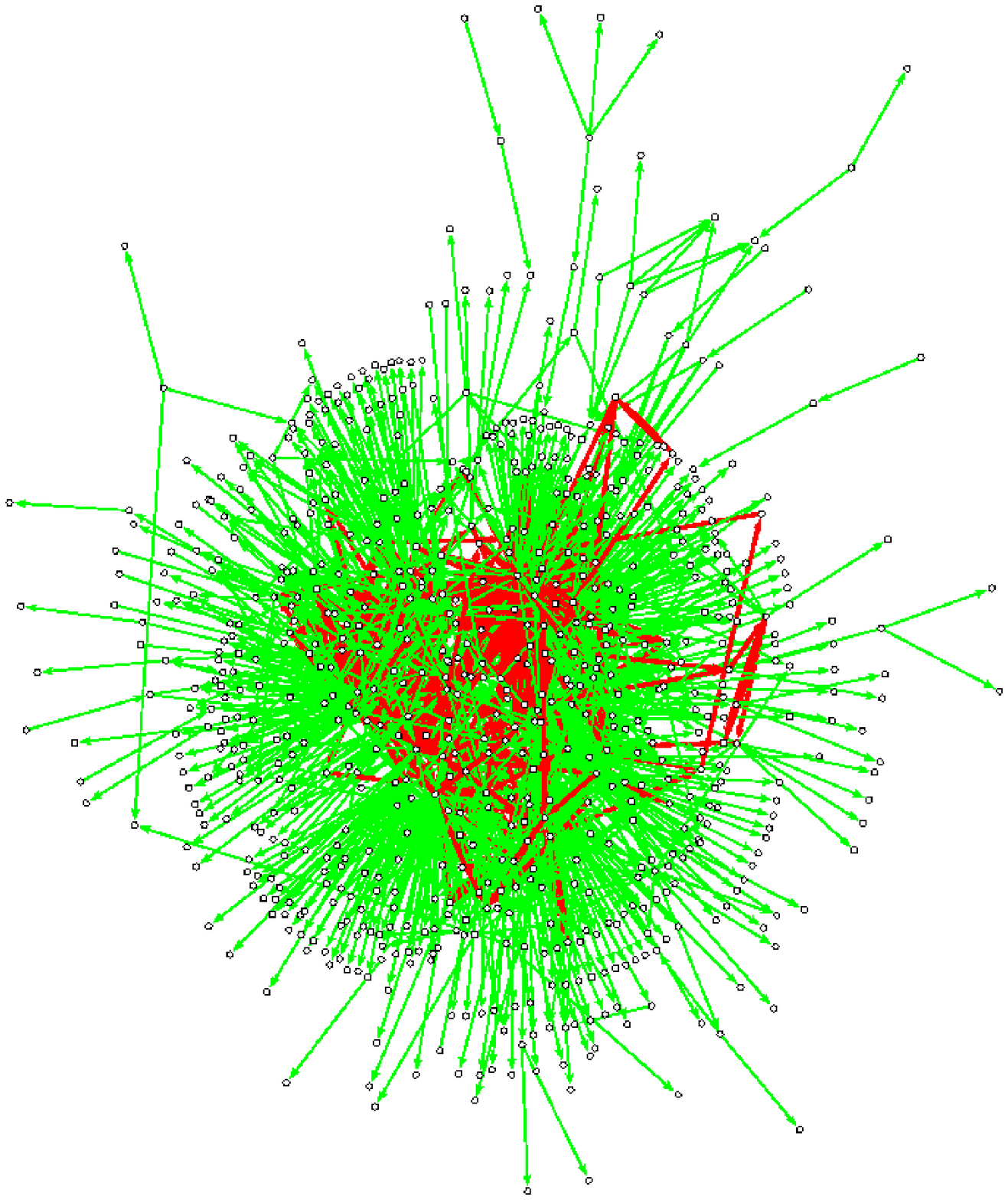}}
\centerline{\includegraphics[width=3in,angle=0]{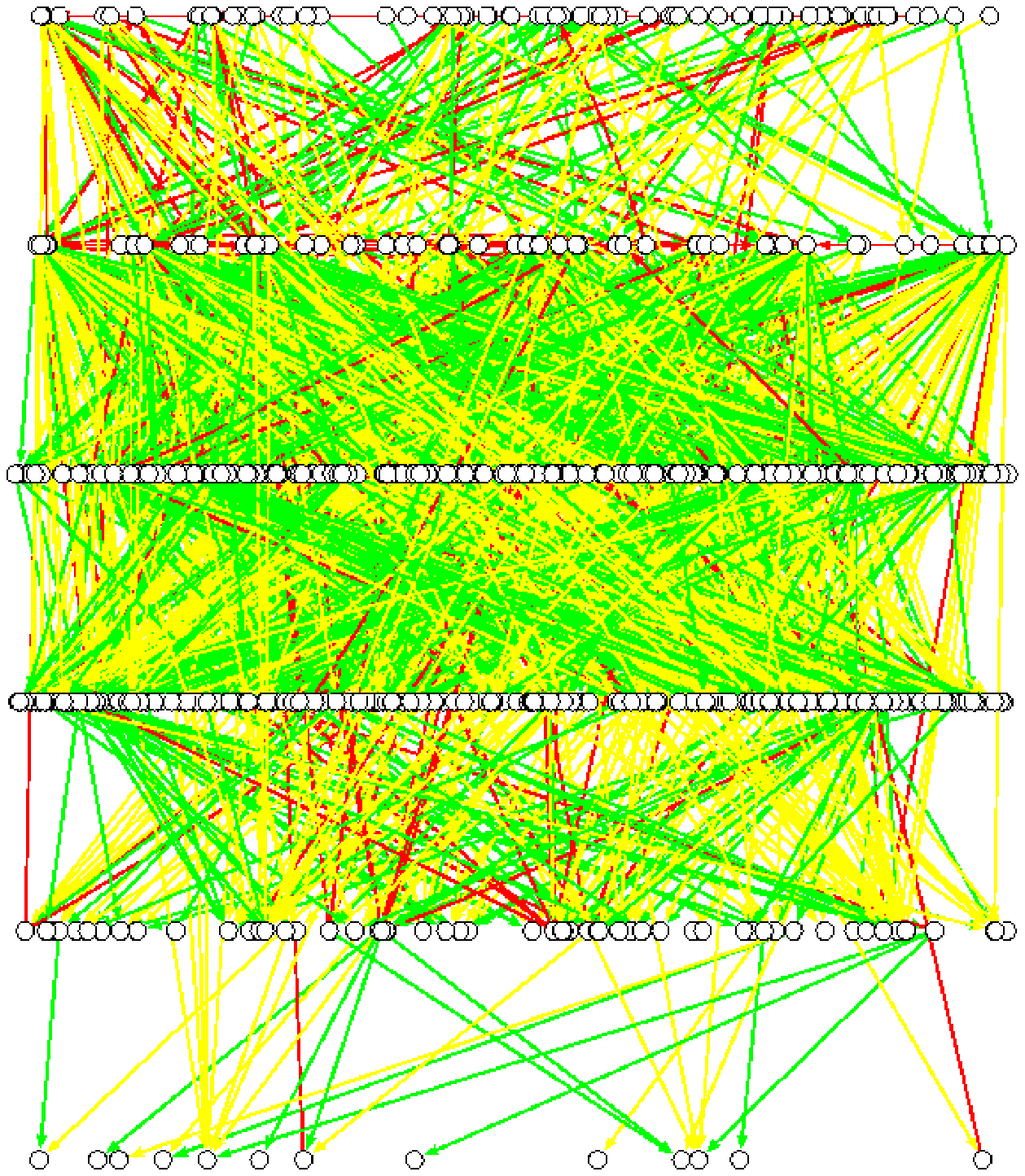}}
\caption{Caption, A part of the post-translational regulatory network in
human shown here includes 1671 automatically and manually curated
protein modification interactions (phosphorylation, proteolytic cleavage,
etc.) between 732 proteins from our ResNet database
\citealp{Resnet}.
Panel A contains the ``hairball'' visualization of the
network structure emphasizing interconnections between
individual pathways. Red edges lie within the strongly connected
component of this network consisting of 107 proteins that could
all be linked to each other by a path in both directions. This makes
any two of these proteins to be simultaneously upstream and downstream
from each other. In Panel B we optimally distribute
proteins over a number of hierarchical levels.
Red arrows represent 208 putative feedback links going from lower
levels of the hierarchy to higher ones, while yellow
ones -- 512 feed-forward links jumping over one or more
hierarchical levels.
Only proteins and links reachable from one of the
71 receptors placed at the top hierarchical level
were included.
}\label{fig:01}
\end{figure}

During the last several years, a substantial amount of
information on large-scale structure of intracellular regulatory
networks has been accumulated.
However, the growth in our understanding of how these networks manage to
function in a robust and specific manner was lagging
behind the shear rate of data acquisition. The fact these
networks are frequently visualized as a giant ``hairball'' (Fig. \ref{fig:01})
consisting of a multitude of edges, linking most constituent
protein-nodes to each other serves as a striking illustration of
the complexity of the issue at hand.

To understand the functioning or even to efficiently
visualize a densely interconnected directed  network it is
desirable to determine the dominant direction of information flow
and to identify links that go against this flow and thus close feedback loops.
Ordering a network with respect to the dominant direction of information flow
can help to determine its previously unknown inputs and
outputs, to
track back hidden sources of perturbations based
on their observable downstream effects, etc.
A simple-minded hierarchical layout of a densely interconnected
network is often impossible due to a ubiquitous presence of feedback loops.
Indeed, all nodes in a strongly connected component of a network
by definition are simultaneously upstream and downstream of each
other. However, if most feedback loops are closed by relatively
few feedback signaling links, the dominant direction of information flow could
still be reconstructed based on a network topology alone.
An identification and removal of these relatively infrequent feedback
links would enable one to perform a hierarchical layout of the remaining
acyclic network which still sufficiently resembles the original one.

In this work we consider the problem of identifying the minimum set of
links, removal of which would render a graph acyclic. In the next section we
introduce two rather different algorithms allowing one to approximately
accomplish this goal, a deterministic 'greedy'
algorithm and a probabilistic Metropolis annealing,
and compare their performance. We find that the probabilistic algorithm
outperforms the deterministic one in better minimizing the number of removed
links, and memory requirements, while maximizing the speed.
A simple visual example is provided for
the situation when the deterministic algorithm is non-optimal.
Following that, we discuss biological implications and applications of our
findings as well as how additional constraints such as {\it a priori}
knowledge of the function and therefore hierarchical position of certain nodes
may affect the algorithm performance.

\section{Approach}

Consider a graph of $N$ vertices labeled as $1, 2, 3, \ldots, N$ and
$L$ directed links labeled by pairs of vertices they connect,
$l_i \equiv (n_i, m_i)$. The goal is to remove as few as possible of the
links to make the graph acyclic, or feedback-free.

An exact
way to solve this problem is to sample all possible combinations
of links to be removed, starting with enumerating individual links,
then pairs of links, etc, until the first acyclic graph is obtained.
Evidently, if a removal of $l$ links finally yields an acyclic graph,
such sampling would require checking the
$\sum_{i=1}^l \binom{L}{i}$ networks for cycles.
For the biologically relevant values
of $L\sim 10^3 - 10^4$ and $l\sim 10 - 10^2$ this approach is clearly
unfeasible. \footnote{ From an obvious identity, $\sum_{i=1}^{L/2}
\binom{L}{i}=2^{L-1}$, it follows that even for fairly modest $L=10^2$ and
  $l=L/2$ the number of such attempts is $ \sim 10^{15}$.}

\subsection{Greedy algorithm}

A natural reduction of such exact enumeration approach is a ``greedy''
algorithm which performs the  ``steepest descent'' in  the number of cycles.
We implemented the following realization of such link removal algorithm:
\begin{itemize}
\item By enumerating all cycles in a graph, each link is assigned a
score equal to the number of cycles it is a member of.
\item The link with the
highest score is removed. When more than one link have the same
highest score, a link to be removed
is selected among the highest-scored ones by
random.
\item This procedure of cycle enumeration and link removal is repeated until
no cycles  are found.
\end {itemize}

The cycle enumeration can be implemented by following
all paths that originate from a given
vertex and recording only the cycles that
come back to this vertex. The procedure is repeated for each of the $N$ graph
vertices: evidently, each cycle of length $C$ is counted $C$ times and a
proper normalization is performed.

\begin{figure}[!tpb]
\centerline{\includegraphics[width=2in,angle=0]{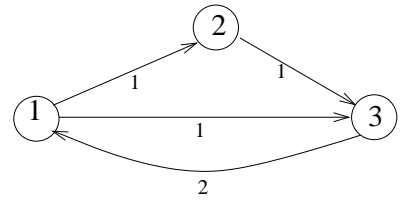}}
\caption{
Caption, Removal of a single $(3,1)$ link makes this 3-vertex graph
acyclic.
}\label{fig:02}
\end{figure}

An example of network where the greedy algorithm performs flawlessly
is shown in Figure \ref{fig:02}.

\begin{figure}[!tpb]
\centerline{\includegraphics[width=3in,angle=0]{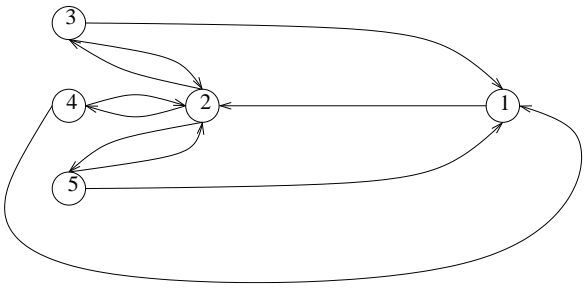}}
\caption{Caption, An example of network where the greedy algorithm fails to
  determine the optimal solution. The link $(1,2)$ carries the highest score 3
  and thus is cut first. However, three 2-node cycles
  $\{2,3\}$,  $\{2,4\}$, and   $\{2,5\}$ remain to be eliminated, after which
  the number of removed links becomes 4. The optimal solution would
  be to cut only three links $(2,3)$, $(2,4)$, and $(2,5)$, each carrying the
  score 2. This optimal solution has almost always been found by the annealing
  algorithm.
}\label{fig:03}
\end{figure}

Here the link $(3,1)$ carries the maximum score 2. A removal of
this link indeed makes the graph acyclic, while a removal of any other than
$(3,1)$ link would require a subsequent removal of the second link to achieve
the same goal. However, one would
suspect that as any ``steepest descent'' method, the proposed greedy
algorithm, performing a sometimes near-sighted local one-step optimization,
may miss the globally optimal solution. This is indeed often the case for
bigger
and more complex graphs; a fairly simple example is given in
Fig. \ref{fig:03}.

\subsection{Simulated annealing network ordering}

\begin{figure}[!tpb]
\centerline{\includegraphics[width=3in,angle=0]{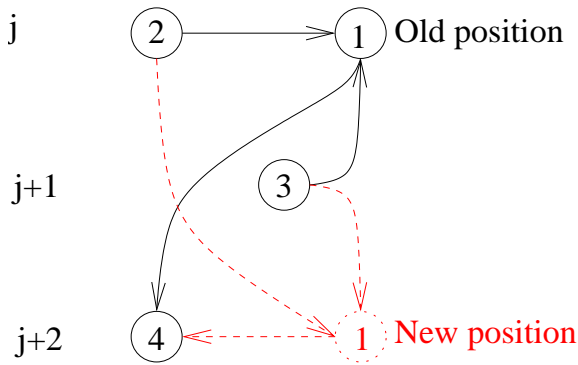}}
\caption{Caption, Node 1 with two incoming and one outcoming link is
selected to move from its current position on
level $j$ to a new position on level $j+2$. The associate energy difference is
$\Delta E = -1 -1 +1 = -1$ where  two -1 contributions come from making
$(2,1)$ and $(3,1)$ links hierarchical and the single +1 contribution comes
from turning the link $(1,4)$ from hierarchical to non-hierarchical.
}\label{fig:03d}
\end{figure}

The task of finding the minimum number of links, cutting which makes the
graph acyclic, can be interpreted as an optimization problem and
tackled by probabilistic methods such as simulated annealing.
Evidently, there exist more than one way to define the optimization function,
and after exploring several possibilities we converged to the following one:
\begin{itemize}
\item For a given network,
a set of $M$ levels is introduced ($M\leq N$, in reality, $M\ll N$ and is of
the order of the graph diameter).
Initially, all nodes are distributed on the levels randomly.
\item For a particular distribution of nodes on levels, the
number of links
that go opposite to the hierarchy, that is, from a lower level
to the same or a higher one, is declared to be the  energy $E$ of the
distribution,  or the
optimization function.
\item A node and its new level are selected at random. A difference in energy
$\Delta E$
that would occur if the node were moved to the new level is calculated. The
node is moved to this new level with the probability $\min\{1, \exp (-\Delta
E/T\}$, where $T$ is the temperature.
\item After the network has been sampled a sufficient number of times (of the
  order of $N \times M$), the temperature is reduced by some factor, usually
  0.9. Initially, the temperature is set sufficiently high, usually of the
  order of the average node degree $L/N$, to allow un-obstructed level
  changes.
\item When the temperature becomes low enough to inhibit any level changes,
  the remaining ascending and in-level links are declared feedbacks and
  removed.
\item The whole procedure can be repeated several times to check for
  consistency in the assignment of feedback links and to determine the
  lowest in the number of removed links solution.
\end{itemize}
A change of level event and the associated energy difference
is illustrated in Fig. \ref{fig:03d}

\begin{figure}[!tpb]
\centerline{\includegraphics[width=3in,angle=0]{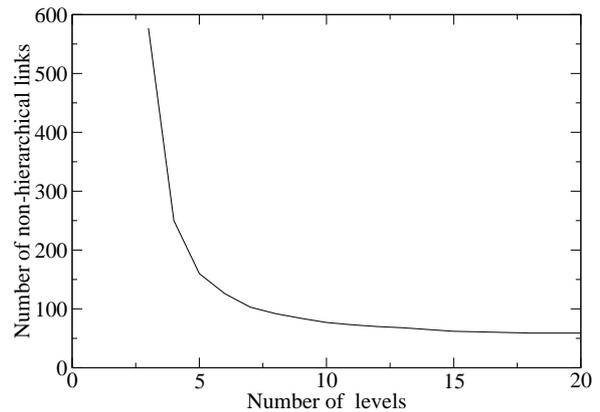}}
\caption{Caption, The number of non-hierarchical links vs the number of levels
$M$  in the annealing layout of the combined (a union of \citealp{Peri2003}
and \citealp{Resnet} datasets) protein phosphorylation network in human cell.
The network consists of $L=2880$ links and $N=1297$ nodes (proteins).
The nodes with zero in-degree and zero out-degree are always put on the top
and bottom levels, correspondingly. The leftmost data point corresponds to the
single intermediate level (3 levels total), the number of non-hierarchical
links clearly reaches its minimum of 59 links for $M\ge 18$.
}\label{fig:04}
\end{figure}

A useful property of this algorithm is that in addition to making a network
acyclic, it also produces a hierarchical layout. The
number of levels $M$ could be fixed by the requirements for such layout.
Otherwise, $M$ could be determined self-consistently, by observing when the
number of counter-hierarchical links stops to decrease upon the increase in
the number of levels. This is illustrated in Fig. \ref{fig:04}
where  a plot of the
number of non-hierarchical links vs number of levels is presented for the
human protein phosphorylation network.

\section{Discussion}
 In the previous section we introduced two algorithms intended to make a
 network acyclic by removing the least number of links.  The stochastic
 stimulated annealing level-ordering algorithm outperforms the deterministic
 greedy algorithm in all respects. Indeed, the greedy algorithm requires
 tracking along all paths originating from a given vertex, which uses a lot of
 memory  and slows the performance significantly. We found it impractical to
 apply the greedy algorithm to networks with more than 100 -- 200
 vertices. This rules out its use for all-organism network ordering and limits
 its utility to analyzing isolated systems and pathways. In addition, as we
 also showed above, it often fails to find the optimum solution, while the
 properly executed stimulated annealing always has a certain probability of
 converging to it. That said,
 there is a grain of biological utility in the ability to
 determine how many cycles pass through a given link. Indeed, the demands for
 robustness 
 in evolution of bio-molecular networks may have resulted in a
 vast redundancy of pathways sending signals along the dominant
 direction of information flow and thus in a relative scarcity of
 links going in the opposite direction. Many of these
 ``backwards'' links simultaneously close up multiple feedback loops.
 The identification of such highly universal feedback links is
 facilitated  by the first, cycle counting stage of the greedy
 algorithm.

Often there exist some {\it a priori} knowledge on the hierarchical
positions of certain network nodes. For example, many of the
receptor proteins localized in the membrane upon activation pass the signals downstream
signaling cascades made of proteins localized in the cytoplasm and ultimately in the cell's
nucleus. Thus receptor proteins might have to be forcefully put on the upper levels of the
hierarchical layout of such signaling network. Contrary
to receptors, many transcription factors serve the role of effectors of
signaling pathways and thus must occupy the lowest levels of the hierarchy.
Initial, or possibly permanent, position of such nodes on the hierarchical
levels often helps to converge to the better in terms of fewer feedback links,
or more biologically relevant solution.

In a similar way, the orientation of certain links (or equivalently, pairs of
nodes) could be quenched if they are known to be of the feed forward of
feed back nature. Based on the initial knowledge of network functioning,
it is also possible to assign a certain weight to a link, so that the energy
$E$ of a particular assignment of nodes to layers is a sum of weights of the
counter-hierarchical links. Thus the {\it a priori} known plausibility of a
link to be (or not to be) a feedback can be introduced into the layering
algorithm.

It is also possible to improve the visual perception of the layout by
shortening the hierarchical links. In its present edition, a ``good'' or
hierarchical link may be arbitrary long, i.e. go down many levels, without
carrying any energetic penalty. This interferes with identifying the
hierarchical levels as certain stages of network flow. Introduction of a small
energetic penalty for particularly long links may alleviate this shortcoming.

We leave these questions as well as those of particular  application of ordering
algorithms to catalytic signaling and transcription regulation cellular
networks for future studies and publications.








\section*{Acknowledgement}
This work was supported by 1 R01 GM068954-01 grant from the NIGMS.
Work at Brookhaven National Laboratory was carried out under
Contract No. DE-AC02-98CH10886, Division of Material Science, U.S.
Department of Energy.
II thanks Theory Institute for Strongly Correlated and
Complex Systems at BNL for financial support during his
visits.


\begin{thebibliography}{}

\bibitem[Nikitin {\it et~al}., 2003]{Resnet}
Nikitin, A., et al (2003) Pathway studio - the analysis and navigation of
molecular networks {\it Bioinformatics} {\bf 19}, 1-3.

\bibitem[Peri {\it et~al}., 2003]{Peri2003} Peri, S. et al. (2003) Development
  of human protein reference database as an initial platform for approaching
  systems biology in humans. {\it Genome Research} {\bf 13}, 2363-2371.

\end{thebibliography}
\end{document}